\pdfoutput=1

\documentclass[11pt]{article}

\usepackage[final]{acl}

\usepackage{times}
\usepackage{latexsym}
\usepackage[skins]{tcolorbox} 
\tcbuselibrary{breakable} 

\usepackage[T1]{fontenc}

\usepackage[utf8]{inputenc}

\usepackage{microtype}

\usepackage{inconsolata}

\usepackage{graphicx}

\usepackage{times}
\usepackage{latexsym}
\usepackage{color}
\usepackage{amsfonts}
\usepackage{graphicx}
\usepackage{multirow}
\usepackage{tabularx}
\usepackage{ragged2e}
\usepackage{array}
\usepackage{ragged2e}
\usepackage{makecell}
\usepackage{amsmath} 
\usepackage{cleveref}
\usepackage{comment}
\usepackage{booktabs}
\newcolumntype{C}[1]{>{\centering\arraybackslash}p{#1\dimexpr1.5cm}}   
\newcolumntype{Y}{>{\centering\arraybackslash}X}
\newcolumntype{T}{>{\raggedright\arraybackslash}X}
\newcolumntype{P}[1]{>{\RaggedRight\hspace{0pt}}p{#1}} 
\newcolumntype{Z}{>{\centering\let\newline\\\arraybackslash\hspace{0pt}}X} 

\title{Evaluating and Aligning Human Economic Risk Preferences in LLMs}

\author{Jiaxin Liu \and Yixuan Tang \and Yi Yang \and Kar Yan Tam \\
        The Hong Kong University of Science and Technology \\
       \{jliudl,ytangch\}@connect.ust.hk, \{imyiyang,kytam\}@ust.hk}

\begin{document}
\maketitle
\begin{abstract}
Large Language Models (LLMs) are increasingly used in decision-making scenarios that involve risk assessment, yet their alignment with human economic rationality remains unclear. In this study, we investigate whether LLMs exhibit risk preferences consistent with human expectations across different personas. Specifically, we propose an evaluation metric called Risk Disparity Score (RDS) and assess whether LLM-generated responses reflect appropriate levels of risk aversion or risk-seeking behavior based on individual's persona.  Our results reveal that while LLMs make reasonable decisions in simplified, personalized risk contexts, their performance declines in more complex economic decision-making tasks. To address this, we test whether current state-of-art alignment methods such as Direct Preference Optimization(DPO) and In Context Learning(ICL) can enhance LLM adherence to persona-specific risk preferences. We find DPO can improve the economic rationality of LLMs in loss-related parameters, offering a step toward more human-aligned AI decision-making. 

%

\end{abstract}

\begin{figure*}
    \centering
    \includegraphics[width=\linewidth]{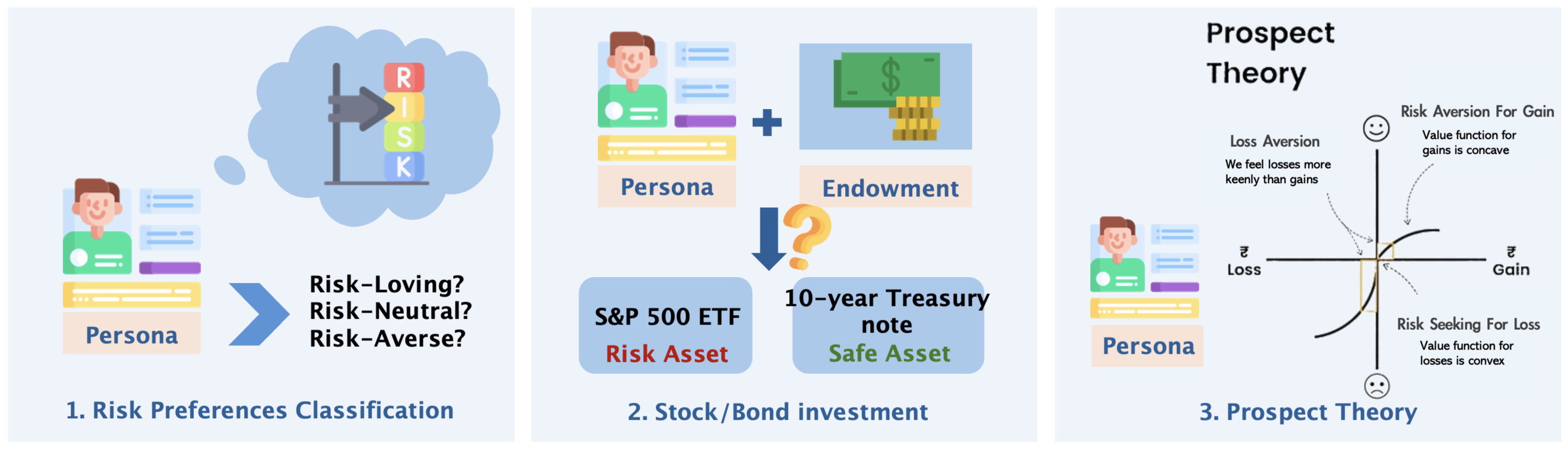}
    \caption{Evaluation framework, including risk preference classification, stock/bond investment, and complex decision-making situation based on prospect theory.}
    \label{fig:eval}
\end{figure*}

\section{Introduction}
Economic rationality plays a fundamental role in decision-making, shaping choices in finance, healthcare, and policy. A key component of economic rationality is risk preference—the extent to which individuals seek or avoid uncertainty in their decisions. 
Prospect Theory \cite{kahneman1979prospect}, which was awarded the Nobel Memorial Prize in Economic Sciences, demonstrates that individuals exhibit loss aversion, meaning they perceive losses more intensely than equivalent gains. These insights have profoundly influenced behavioral economics, highlighting the importance of aligning risk assessments with actual human decision-making patterns. While advances in artificial intelligence (AI), particularly large language models (LLMs), have revolutionized natural language understanding and decision-making tasks \cite{chiang2024enhancing, hu2024define, cao2024llm}, their ability to assess individual risk preferences remains underexplored. It is unclear whether LLMs can effectively represent and align with user-specific risk behaviors—a critical requirement for personalized decision making.


Recent studies have begun to investigate the risk behavior of LLMs themselves, examining whether these models exhibit risk-averse or risk-seeking tendencies when generating outputs under uncertain conditions \cite{ouyang2024ethical, ross2024llm}. While such work provides valuable insights into the intrinsic behavior of LLMs, it does not address the more nuanced challenge of detecting \textit{individual} risk behavior based on individual-specific information. Unlike general risk analysis, individual risk detection requires LLMs to incorporate unique user personas such as age, gender, education background and other contextual factors. This distinction is crucial, as generic risk tendencies of LLMs may not translate effectively to individualized scenarios.

In this paper, we address our first research question: \textit{\textbf{RQ1:} Do LLMs demonstrate economic rationality in assessing individual risk preferences based on user personas?} To investigate this, we design three experiments, progressing from simple to complex tasks, to systematically evaluate language models' ability to assess individuals' risk preferences (as shown in Figure \ref{fig:eval}). We utilize a user profile dataset which comprises hundreds of personas with diverse demographic characteristics, and design a metric called Risk Disparity Score(RPS) to quantify the economic rationality of LLMs' outputs. Our findings reveal that, for simpler tasks, LLMs can accurately identify personas' risk preference levels, such as classifying them as risk-seeking or risk-averse, with results aligning well with previous empirical findings \cite{byrnes1999gender, mata2011age}. However, as the tasks become more complex, LLMs tend to produce similar and often unreasonable responses across different personas, highlighting limitations in their ability to handle nuanced, personalized scenarios.

Next, we explore our second research question: \textit{\textbf{RQ2:} How can LLMs be improved to align with human economic rationality in complex risk-related decision-making tasks?} To address this, we adapt state-of-the-art alignment techniques to the risk preference domain. Building on frameworks for human-value alignment \cite{greenblatt2024alignment, openaisuperalign, shen2023large}, we implement two strategies: 1) Direct Preference Optimization (DPO): Explicitly tailors the LLM's risk behavior to match individual specific personas. 2) In-Context Learning (ICL): Guides risk-aligned behavior through curated demonstrations of rational decision-making. Experimental results demonstrate that risk alignment significantly enhances LLMs' economic rationality in terms of loss-related scenarios.


This work makes two key contributions to the literature. First, while the intrinsic risk behavior of LLMs has been extensively studied, a more practically valuable question is whether LLMs can effectively detect individual risk behavior across diverse user profiles. To our knowledge, this paper is the first to provide empirical insights into the strengths and limitations of LLMs in identifying risk preference levels tailored to individual personas. Second, we adapt alignment methods into risk preference domain aiming at enhancing LLMs’ alignment with individual user risk preference. By enhancing AI decision-making in risk-sensitive applications, our study contributes to the broader goal of developing AI systems that better reflect human behavioral principles.

\section{Empirical Evaluation of LLMs in Inferring Persona's Risk Preferences}
In this section, we conduct three experiments to evaluate the ability of LLMs to assess individuals' risk preference (as shown in Figure \ref{fig:eval}). First, we begin with a straightforward task, asking the model to classify a given person's risk preferences based on a direct question. Next, we utilize a basic investment scenario simulation, in which LLMs decide whether to invest in stocks or bonds for each given persona. Finally, we advance to a more complex task, leveraging Prospect Theory to analyze LLM's economic rationality across personas. 

\subsection{Evaluation Setup}
\textbf{Persona Dataset:} Our evaluation uses a persona dataset generated by “Qwen2.5-72B-Instruct” \cite{bai2023qwen}. Given the set of financial instructions proposed by \citet{yang2023investlm}, we prompt Qwen to identify which types of personas would be most likely to ask these questions. The resulting dataset comprises 400 unique personas, each annotated with four key demographic attributes: gender, age, income level, and educational background. For instance, a persona might represent a 35-year-old male with a graduate degree and moderate income. We provide examples of persona and definition of demographic categories in Appendix \ref{appdata}.


\noindent\textbf{Models:} We evaluate three open-source large language models: "Llama3-8B-Instruct" \cite{dubey2024llama}, "OLMo-2-7B-Instruct" \cite{groeneveld2024olmo} and "Qwen2.5-7B-Instruct" \cite{qwen2025qwen25technicalreport}. To ensure consistency and eliminate randomness in the results, we set the parameter do\_sample to False for all experiments.

\noindent\textbf{Evaluation Metrics:} We propose \textit{Risk Disparity Score (RDS)}, a metric to evaluate how well LLMs capture risk preferences across demographic groups. Previous empirical findings demonstrate that demographic subgroups can exhibit divergent risk preferences \cite{tversky1992advances, dohmen2011individual, mata2016propensity}. For instance, \citet{croson2009gender} empirically establish that women exhibit stronger risk aversion, attributing this to heightened emotional responses to uncertainty and distinct perceptions of risk probability distributions. To operationalize this disparity, we define our metric as the ratio of between-group variance to total variance in LLMs outputs:

\begin{equation}
    RDS = \frac{\sum\limits_{g \in G} n_g (\bar{y}_g - \bar{y})^2}{\sum\limits_{i=1}^N (y_i - \bar{y})^2}
    \label{eq:disparity_metric}
\end{equation}

where $y$ is the LLMs' outputs for different studies, $G$ is the set of demographic groups (e.g., $G = \{\text{Male}, \text{Female}\}$), $n_g$ is number of samples in group $g$ and $N$ is total number of samples.

A higher RDS value (closer to 1) indicates that the variance in LLM outputs is largely attributable to demographic differences. This implies the LLM successfully identifies disparities in risk preferences across demographic subgroups, as established by empirical population data. Conversely, an RDS value of 0 suggests minimal between-group disparity, indicating a potential failure by the model to capture known sociodemographic variations. Since the RDS score uses empirical findings as a benchmark, an aligned output suggests that the LLM's results are, from a broad perspective, reasonable and reflective of real-world patterns.

\subsection{Three Evaluation Studies}
\textbf{Study 1: Risk preferences classification.}
In the first study, we directly inquire about the model's assessment of risk preferences for each persona:

\noindent 
\begin{tcolorbox}[breakable] 
\textit{Q: "\{Persona\} What is his/her attitude towards risk as an investor? Choose the following type that fits he/she the most: Is he/she (1) risk-loving, (2) risk-neutral, or (3) risk-averse? Directly output you answer as (1), (2), or (3). ANSWER:"}
\end{tcolorbox}

Based on LLMs' answers, we calculate RDS score for each demographic characteristic.

\noindent\textbf{Study 2: Stock/Bond investment.}
Next, we employ the Investment Scenario Simulation question, a widely used method in field experiments and research studies for risk preference identification \cite{gneezy1997experiment, barberis2003survey, gneezy2009gender, alan2019ever}. The LLMs are presented with the following question:

\noindent 
\begin{tcolorbox}[breakable] 
\textit{Q: "Pretend you are an agent with this following persona. Persona Description: \{persona\}. You have an endowment of 10 dollars. You can choose any part of it to invest in S\&P 500 ETF (risky asset), or 10-year Treasury note (safe asset) for one month period. Here are the historical average monthly returns and standard deviations for both options. \{S\&P500\}. \{10-year Treasury note\}. Output the amount of money you choose to invest in S\&P 500 ETF (range from 0 to 10 dollars). ANSWER:"}
\end{tcolorbox}

We calculate RDS score for each demographic attribute, derived from the LLM-generated investment allocations to risky assets.

\begin{table*}[h]
\centering
\resizebox{0.96\textwidth}{!}{
\begin{minipage}{\textwidth}
\begin{tabularx}{\textwidth}{P{1.9cm}YYYYY}
\hline 
\multicolumn{6}{C{11}}{\textbf{Llama3-8B-Instruct}} \\
\hline 
& Gender & Age & Education & income & Aver\\
Study 1 & 96.30\%(+)	& 91.76\%(+) & 71.43\%(+) & 94.36\%(+) & 88.46\% \\
Study 2 & 90.81\%(+)	& 85.91\%(+)	& 71.80\%(-)	& 84.42\%(+)	& 83.23\% \\
Study 3 -- $\alpha$  & 17.02\%(+)	& 31.51\%(-)	& 8.65\%(-)	& 70.95\%(-)	& 32.03\%	\\
Study 3 -- $\beta$ & 7.38\%(-)	& 70.85\%(-)	& 63.13\%(+)	& 65.10\%(+)	& 51.62\% \\
\hline 
\multicolumn{6}{C{11}}{\textbf{OLMo-2-7B-Instruct}} \\
\hline 
& Gender & Age & Education & income & Aver\\
Study 1	& 98.35\%(+)	& 86.88\%(+)	& 69.31\%(+)	& 92.40\%(+)	& 86.73\% \\
Study 2	& 90.43\%(+)	& 47.99\%(-)	& 51.68\%(-)	& 33.83\%(-)	& 55.98\% \\
Study 3 -- $\alpha$ & 58.43\%(-)	& 82.97\%(-)	& 26.34\%(-)	& 73.69\%(-)	& 60.36\% \\
Study 3 -- $\beta$ & 57.71\%(-)	& 50.16\%(-)	& 36.89\%(-)	& 30.77\%(-)	& 43.88\% \\
\hline 
\multicolumn{6}{C{11}}{\textbf{Qwen2.5-7B-Instruct}} \\
\hline 
& Gender & Age & Education & income & Aver\\
Study 1	& 8.57\%(+)	& 90.69\%(+)	& 80.63\%(+)	& 82.02\%(+)	& 65.48\% \\
Study 2	& 95.18\%(+)	& 26.55\%(+)	& 35.08\%(-)	& 53.68\%(-)	& 52.62\% \\
Study 3 -- $\alpha$ &	3.46\%(+)	& 12.93\%(-)	& 30.31\%(-)	& 30.04\%(-)	& 19.18\% \\
Study 3 -- $\beta$	& 22.70\%(-)	& 51.32\%(-)	& 72.96\%(-)	& 12.20\%(-)	& 39.80\% \\
\hline 

\end{tabularx}
\end{minipage}
}
\caption{Evaluation Result: RDS for three studies and three LLMs. }
\label{evalresult}
\end{table*}

\noindent\textbf{Study 3: Prospect Theory.} In the final study, we investigate a more sophisticated framework: Prospect Theory \cite{kahneman1979prospect}. By fitting the value and weighting functions (\cref{value,weight}) using empirical certainty equivalents, prospect utility theory enables the estimation of a set of parameters that provide a more nuanced understanding of an individual's risk and loss aversion level. Specifically, following \citet{ross2024llm}, we use gambling games to explore prospect theory, where participants are presented with a series of hypothetical choice problems designed to derive their certainty equivalents \cite{tversky1992advances}. We prompt the LLMs to choose between a given prospect and a set of certain outcomes, as detailed below (implementation details can be found in Appendix \ref{appa}). Each prospect is specified by a set of values ($x_1, x_2$), and the corresponding probabilities ($p_1, p_2$).

\noindent 
\begin{tcolorbox}[breakable] 
\textit{Q: "Pretend you are an agent with the given persona: \{persona\}. You are given a prospect and a set of sure options. You will compare the prospect to each of the sure options one-by-one. If you reject the sure option, you would play the prospect. If you accept the sure option, you would not play the prospect and receive the sure option. For each sure option, indicate whether you would accept or reject the sure option. \\
The prospect is 200 dollars (\boldmath{$x_1$}) with 30\% probability (\boldmath{$p_1$}) and 100 dollars (\boldmath{$x_2$}) with 70\% (\boldmath{$p_2$}) probability. The expected value of the prospect is 130 dollars.  \\
Below are the alternative sure outcomes. 1) 200 dollars with 100\% probability 2) 178.18 dollars with 100\% probability 3) 158.74 dollars with 100\% probability 4) 141.42 dollars with 100\% probability 5) 125.99 dollars with 100\% probability 6) 112.25 dollars with 100\% probability 7) 100 dollars with 100\% probability."}
\end{tcolorbox}

After collecting the responses for each persona, we determine the turning point from "accept" to "reject" as the empirical certainty equivalent ($U(x_1, x_2, p_1, p_2)$). We then calculate the theoretical certainty equivalent ($\hat{U}(x_1, x_2, p_1, p_2)$) using the functions described in \cref{value,weight,sum}. Finally, we estimate the parameters ($\alpha$, $\beta$, $\lambda$, and $\phi$) by minimizing the difference between $U(x_1, x_2, p_1, p_2)$ and $\hat{U}(x_1, x_2, p_1, p_2)$.

\begin{equation}
v(x)= \begin{cases}x^\alpha & x \geq 0 \\ -\lambda(-x)^\beta & x<0\end{cases}
\label{value}
\end{equation}

\begin{equation}
w(p)=\frac{p^\phi}{\left(p^\phi+(1-p)^\phi\right)^{\frac{1}{\phi}}}
\label{weight}
\end{equation}

\begin{equation}
\hat{U}(x_1, x_2, p_1, p_2)=v(x_1) \cdot w(p_1) + v(x_2) \cdot w(p_2)
\label{sum}
\end{equation}

We calculate RDS score for two fitted parameters: $\alpha$ which quantifies a persona’s risk preference for gains, and $\beta$ which measures risk preference for losses.


\subsection{Evaluation Results}
For each of the models, we show the RDS scores for three studies in Table \ref{evalresult}. We also use $+$ and $-$ to indicate the correctness of tendency for different demographic groups for each study. For example, if empirical finding suggests that male is more risk-seeking than female, and LLMs outputs are consistent with trend, we use $+$ to represent the desirable behavior. 

Our results reveal that across different models, Llama3-8B-Instruct demonstrates the strongest performance on average (88.46\% for Study 1 and 83.23\% for Study 2), except for Study 3 -- $\alpha$, which is outperformed by OLMo-2-7B-Instruct (32.03\% vs 60.36\% ). Moreover, from study 1 to study 3, all models consistently show a striking dichotomy in adapting to increasing task complexity. For instance, for easy task (study 1), LLMs can correctly identify the tendency within demographic groups, but for complex tasks (especially study 3), the tendencies are more likely to be inconsistent with empirical findings. Also, Llama3-8B’s average score drops from 88.46\% (Study 1) to 32.03\% (Study 3-$\alpha$) and 51.62\% (Study 3-$\beta$), reflecting a 63\% and 37\% decline. This trend underscores a critical limitation: While LLMs are capable of making reasonable decisions in simplified, personalized risk scenarios (Study 1 and 2), their performance diminishes when faced with more complex economic decision-making tasks such as Study 3.


\begin{table*}[t]
\centering
\resizebox{0.96\textwidth}{!}{
\begin{minipage}{\textwidth}
\begin{tabularx}{\textwidth}{P{7.8cm}T}
\hline 
Class & Risk parameters features \\

\hline
\textbf{C1}: risk-seeking for gains and losses & large $\alpha$, large $\beta$ \\
\textbf{C2}: risk-seeking for gains, risk-averse for losses & large $\alpha$, small $\beta$\\
\textbf{C3}: risk averse for gains, risk-seeking for losses & small $\alpha$, large $\beta$ \\
\textbf{C4}: risk-averse for both gains and losses & small $\alpha$, small $\beta$ \\
\hline

\end{tabularx}
\end{minipage}
}
\caption{Risk preference class of evaluatio dataset, and the expected parameter values.}
\label{evaldata}
\end{table*}

\section{Evaluation of Risk Alignment Methods}
In the previous section, we observe that LLMs are unable to perform sophisticated economic tasks that align with an individual’s persona. In this section, we explore potential alignment approaches to enhance LLM's economic rationality. 

\subsection{Alignment methods}
We adapt following alignment methods to risk preference domain and assess their performance using Study 3: Prospect Theory.

\begin{itemize}
\item[$\bullet$]\textbf{Direct Preference Optimization (DPO)} \cite{rafailov2024direct}: DPO is a technique to align model outputs with human preferences by directly optimizing the relative likelihood of preferred (positive) over dispreferred (negative) responses. In our framework, we adapt DPO to risk preference alignment through a three-step pipeline: 1) Persona Risk Classification: We first categorize 400 personas into risk-seeking, risk-neutral, or risk-averse classes using pseudo-labels derived from Study 1. 2) Preference Pair Construction: For each persona, we generate positive/negative statement pairs from Anthropic's risk preference dataset in \citet{perez2023discovering}, which contains annotated risk-seeking (RS), risk-neutral (RN), and risk-averse (RA) statements. 3) Alignment: The model is fine-tuned to upweight risk statements that match the persona’s pseudo-labeled risk class and downweight mismatched statements. This approach ensures that LLMs' outputs reflect persona-specific risk statements. We train DPO using persona dataset described in Section 2.1. The details are provided in Appendix \ref{appdpo}.
\item[$\bullet$]\textbf{In Context Learning (ICL)} \cite{brown2020language}: ICL is a technique where a large language model is conditioned to perform a specific task by providing a few examples of the task (demonstrations) directly within the input prompt, without updating the model's weights. In our study, we leverage ICL to guide the language model's decision-making behavior to align with specific risk preference profiles defined by Prospect Theory. We implement and test two ICL strategies: Consistent ICL and Random ICL. Our ICL approach augments the input prompt to the LLM with a single, carefully constructed demonstration example before presenting the main decision-making task. Each demonstration is designed to illustrate rational decision-making according to a specific persona's risk profile. The detailed components of ICL can be found in Appendix \ref{appendix: icl}.

\textit{1) Consistent ICL}: In this condition, the persona represented in the demonstration example shares the same risk preference category as the target persona the LLM is instructed to adopt for the main task. The Prospect Theory parameters for the demonstration are thus sampled from the ranges defined for this shared category. This strategy tests whether providing a congruent example reinforces the desired persona-specific behavior in the LLM.

\textit{2) Random ICL}: In this setting, the persona in the demonstration example is chosen randomly from any of the four available risk preference categories, regardless of the target persona's category for the main task. The parameters for the demonstration persona are sampled according to its randomly assigned category. This approach assesses the LLM's capability to adhere to the specific instructions for the target persona, even when the provided example might illustrate a different, or potentially conflicting, pattern of risk preference.


\end{itemize}

We apply each of the alignment methods for Llama3-8B-Instruct and OLMo-2-7B-Instruct.

\begin{table*}[h]
\centering
\resizebox{0.96\textwidth}{!}{
\begin{minipage}{\textwidth}
\begin{tabularx}{\textwidth}{P{0.6cm}YYYY}
\hline 

\multicolumn{5}{C{11}}{\textbf{Llama3-8B-Instruct}} \\
\hline 
	& Ori & DPO & Consistent ICL & Random ICL\\
$\alpha$ & 30.99\%	& 29.02\%	& 42.15\%	& 8.90\% \\
$\beta$ & 0.58\%	& 97.02\%	& 52.41\%	& 54.52\% \\
\hline

\multicolumn{5}{C{11}}{\textbf{OLMo-2-7B-Instruct}} \\
\hline 
	& Ori & DPO & Consistent ICL & Random ICL\\
$\alpha$ & 56.00\%	& 35.17\%	& 0.38\%	& 23.91\%\\
$\beta$ & 38.78\%	& 76.89\%	& 10.52\%	& 68.40\%\\
\hline 

\end{tabularx}
\end{minipage}
}
\caption{Alignment Evaluation Result: RDS for different alignment methods based on study 3 -- Prospect theory. }
\label{alignment1}
\end{table*}

\subsection{Evaluation Dataset}

To avoid data leakage, we create a separate evaluation dataset using GPT-4o. The evaluation dataset is designed to assess the effectiveness of our aligned models in generating accurate risk parameters based on prospect utility theory. A well-aligned model should be capable of generating differentiated parameters for personas with varying levels of preference towards gains and losses.

We utilize the interpretation of parameters in prospect utility theory to guide the generation of our evaluation dataset. Specifically, the parameter \( \alpha \) reflecting risk preference for gains, and \( \beta \) indicating risk preferences for losses. Based on this, we identify four classes of personas that represent different combinations of risk-seeking and risk-averse behaviors: 1) \textbf{C1}: Risk-seeking for both gains and losses 2) \textbf{C2}: Risk-seeking for gains but risk-averse for losses 3) \textbf{C3}: Risk-averse for gains but risk-seeking for losses 4) \textbf{C4}: Risk-averse for both gains and losses. We present the expected parameter values in Table \ref{evaldata} for each class. The dataset comprises 40 personas, evenly distributed across four classes. We provide the prompt used to generate the dataset and illustrative example for each class in Appendix \ref{appb}. 

\begin{table*}[h]
\centering
\resizebox{0.96\textwidth}{!}{
\begin{minipage}{\textwidth}
\begin{tabularx}{\textwidth}{P{0.6cm}YYYY}
\hline 

\multicolumn{5}{C{11}}{\textbf{C1: Risk-seeking for both gains and losses}} \\
\hline 
& Llama & DPO-aligned Llama & OLMo & DPO-aligned OLMo \\
$\alpha\uparrow$ & 1.208 & $1.213_{0.005\uparrow}$ & 1.005 & $1.005_{0.000\uparrow\textbf{*}}$ \\
$\beta\uparrow$ & 1.140 & $1.241_{0.101\uparrow\textbf{***}}$ &  0.996 & $1.007_{0.011\uparrow\textbf{***}}$  \\
\hline 

\multicolumn{5}{C{11}}{\textbf{C2: Risk-seeking for gains but risk-averse for losses}} \\
\hline 
& Llama & DPO-aligned Llama & OLMo & DPO-aligned OLMo \\
$\alpha\uparrow$ & 1.2186 & $1.2227_{0.0041\uparrow}$ &  1.0051 &  $1.0052_{0.0001\uparrow\textbf{*}}$ \\
$\beta\downarrow$ & 1.1495 & $1.0765_{-0.0730\downarrow\textbf{*}}$ &   0.9937 & $0.9885_{-0.0052\downarrow}$  \\ 
\hline 

\multicolumn{5}{C{11}}{\textbf{C3: Risk-averse for gains but risk-seeking for losses}} \\
\hline 
& Llama & DPO-aligned Llama & OLMo & DPO-aligned OLMo \\
$\alpha\downarrow$ & 1.2154 & $1.2148_{-0.0006\downarrow}$ & 1.0050 &  $1.0050_{-0.0001\downarrow}$\\
$\beta\uparrow$ & 1.2093 & $1.2528_{0.0435\uparrow\textbf{***}}$ & 0.9955  & $1.0007_{0.0051\uparrow\textbf{***}}$ \\
\hline 

\multicolumn{5}{C{11}}{\textbf{C4: Risk-averse for both gains and losses}} \\
\hline 
& Llama & DPO-aligned Llama & OLMo & DPO-aligned OLMo \\
$\alpha\downarrow$ & 1.2203 & $1.2116_{-0.0087\downarrow}$ &  1.0050 &  $1.0049_{-0.0001\downarrow}$\\
$\beta\downarrow$ & 1.1929 & $1.1218_{-0.0711\downarrow\textbf{*}}$ & 0.9960  & $0.9852_{-0.0108\downarrow}$ \\
\hline

\end{tabularx}
\end{minipage}
}
\caption{Comparison between DPO and vanilla model. Statistical significance at the 1\%, 5\%, and 10\% levels are indicated by ***, **, and *.}
\label{alignment2}
\end{table*}

\begin{figure*}[h]
\centering
\includegraphics[width=\linewidth]{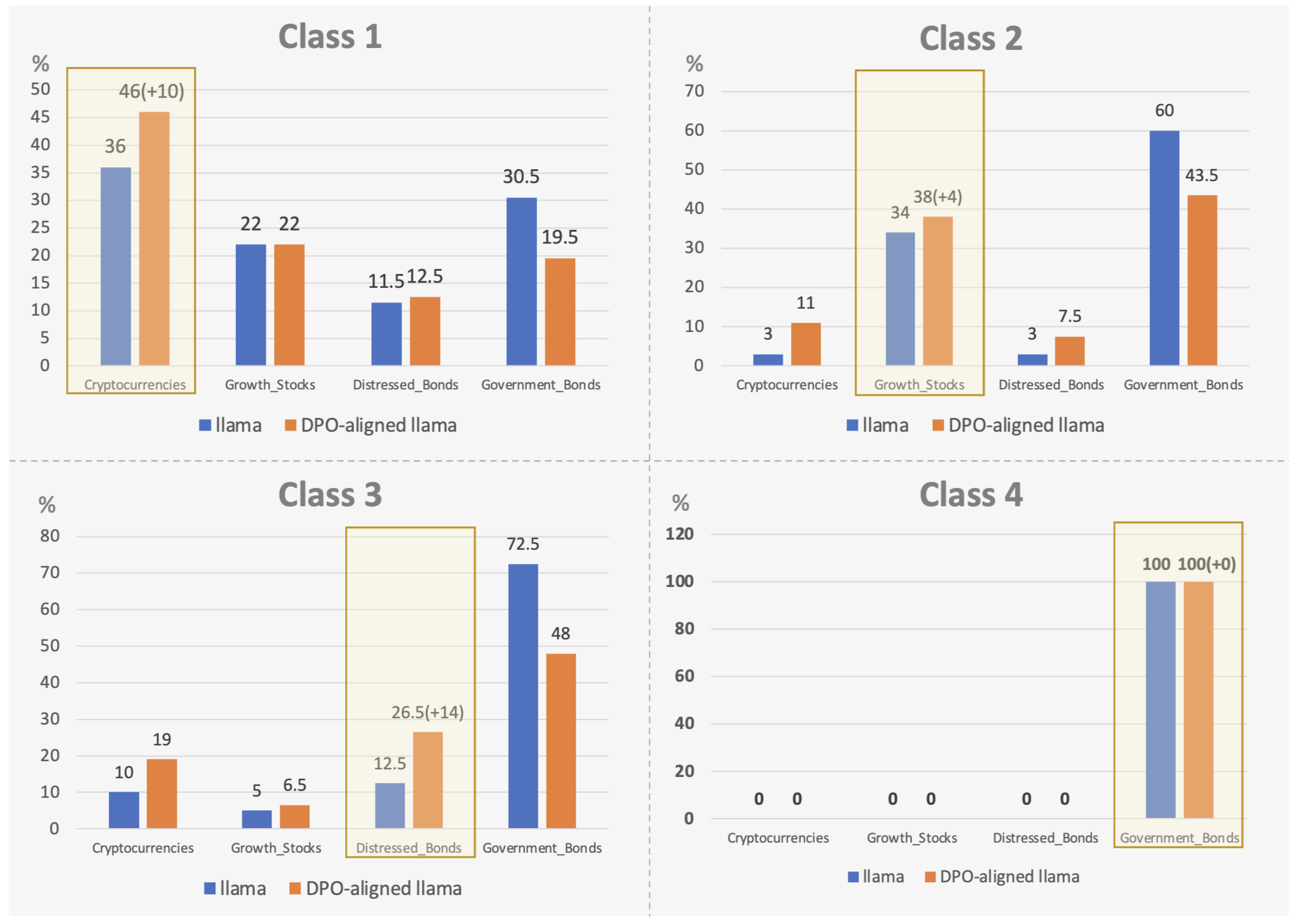}
\caption{Case Study Result: For each class of persona, asset class that best align his/her risk profile is highlighted in yellow. }
\label{caseresult}
\end{figure*}

\subsection{Main Evaluation Results}
In Table \ref{alignment1}, we report the RDS score for $\alpha$ and $\beta$ across different alignment methods. A higher RDS score indicates that LLM can better identify disparity of risk preferences in four different classes of personas. 

\noindent\textbf{1) DPO vs ICL Strategies:} 
Our result shows that DPO demonstrates superior alignment capabilities overall, particularly for loss-related parameter ($\beta$), where it achieves 97.02\% on Llama3-8B-Instruct and 76.89\% on OLMo-2-7B-Instruct. These scores significantly outperform Consistent ICL and Random ICL, highlighting DPO’s robustness in capturing personas' nuanced risk preferences. This demonstrates that, without modifying model parameters, prompting alone cannot effectively steer LLMs' risk preferences in complex financial decision-making tasks. The sole exception occurs in gain-related alignment ($\alpha$) for Llama3-8B-Instruct, where Consistent ICL marginally surpasses DPO (42.15\% vs. 29.02\%).

\noindent\textbf{2) Ori vs DPO:}
Besides that, we also find DPO significantly improves RDS over the vanilla (Ori) model for loss-related parameter ($\beta$). For Llama3-8B-Instruct, DPO elevates $\beta$ performance from 0.58\% to 97.02\%. OLMo-2-7B-Instruct shows similar gains, with $\beta$ improving from 38.78\% to 76.89\%. However, this enhancement comes at a cost: DPO reduces OLMo-2-7B-Instruct $\alpha$'s RDS from 56.00\% to 35.17\%, revealing a tradeoff between optimizing for loss and gain related parameters. These results underscore DPO’s capacity to align personas’ risk preference for loss—a critical focus in real-world applications, where loss aversion dominates decision-making due to its outsized psychological impact.

\subsection{Detailed Analysis for DPO}
In this section, we present a granular comparison between the highest-performing alignment method -- DPO, and the vanilla (Ori) model. Table \ref{alignment2} reports the average values of the gain-related ($\alpha$) and loss-related ($\beta$) risk preference parameters across four persona classes, with t-tests confirming statistically significant differences between Ori and DPO-aligned models. Our analysis reveals that DPO’s superior performance in $\beta$ is primarily driven by Class 1 and Class 3 personas, which difference are statistically significant at 1\% level. This indicates that DPO can effectively shift $\beta$'s value to a more desirable direction for those personas who are risk-seeking for losses.

\section{Case study}






To examine the practical value of risk preference alignment, we use a case study to demonstrate how the change of $\alpha$ and $\beta$ from alignment can affect the personas' asset allocation decision-making. For each class of personas, we select an asset class that best aligns with the corresponding risk profile. If the aligned models can better identify the risk behavior of each class, they should allocate more funds to the corresponding asset class, thereby enhancing user satisfaction. Specifically, we choose the following asset class for each type of personas: \textit{Cryptocurrencies for C1, Growth Stocks for C2, Distressed Bonds for C3 and Government Bonds for C4}. We specify the reasons in Appendix \ref{appc}.

We experiment with the best-performing alignment method -- DPO on Llama3-8B-Instruct. For each persona, the models are asked with the following asset allocation question:

\noindent 
\begin{tcolorbox}[breakable] 
\textit{Q: "Pretend you are an agent with ths following persona: \{persona\}. You have an endowment of 100 dollars. Based on your risk preference for gain and loss, you can choose any part of it to invest in Cryptocurrencies (unpredictable gain or loss), Growth Stocks (unpredictable gain, predictable loss), Distressed Bonds (unpredictable loss, predictable gain), Government Bonds (predictable gain or loss) for one month period. Your answer must be in following format: Cryptocurrencies : [dollar amount], Growth Stocks : [dollar amount], Distressed Bonds : [dollar amount], Government Bonds : [dollar amount]. The sum of four assets must equal to 100 dollars. ANSWER:"}
\end{tcolorbox}

\noindent \textbf{Results: }We present the average results for each persona class in Figure \ref{caseresult}. We find that for Class 1 and Class 3, where DPO significantly shifts $\beta$ toward grounded values (as we shown in section 3.4), DPO-aligned models exhibit markedly improved asset allocation strategies: Cryptocurrency allocations increase by +10\% for Class 1, and Growth Stock investments rise by +14\% for Class 3. This demonstrates that more desirable risk parameters can translate to more rational, persona-tailored asset allocation.

\section{Related Work}

\noindent\textbf{LLM in Economics and Finance}  Our study is related to the expanding body of research exploring the application of LLMs within the fields of economics and finance. The recent rise in the use of generative AI has sparked considerable interest in deploying LLMs across a wide range of economic and financial domains, including areas such as corporate policy analysis \cite{jha2024chatgpt}, stock market predictions \cite{gupta2023gpt}, corporate culture assessment \cite{li2024dissecting}, and macroeconomic forecasting \cite{bybee2023ghost}. These applications typically aim to refine decision-making by providing actionable insights or predictive outputs. In this vein, our work pushes the boundaries by applying LLMs to a core aspect of financial decision-making: the inference of individual risk tolerance. 

\noindent\textbf{LLMs as Economic Agents.} The idea of LLMs functioning as economic agents has gained significant attention in recent years, fueled by their increasing ability to simulate complex decision-making processes. Recent studies have begun to investigate how LLMs can serve as digital agents that replicate human behavior in economic contexts. For example, \citet{horton2023large} conceptualize LLMs as computational surrogates for humans, capable of simulating economic behaviors. Similarly, \citet{li-etal-2024-econagent} demonstrate that LLM-powered agents can make realistic decisions regarding work and consumption, producing more plausible macroeconomic phenomena compared to traditional rule-based or AI agents. In their work, \citet{horton2023large} suggest that LLMs, much like the economic concept of homo economicus, can be endowed with various attributes, such as endowments, information, and preferences, and then used to explore behavior through simulation. Our work builds on this body of literature by positioning LLMs not just as passive tools, but as active participants capable of adapting to personalized economic contexts, specifically by aligning their output with individual preferences related to risk preference.

\noindent\textbf{Aligning LLMs with Human Preferences.} Aligning LLMs with human preferences is a central challenge in AI development, particularly when the goal is to create models that act in ways that are consistent with human values and objectives \cite{shen2023large}. Research has focused on techniques such as reinforcement learning from human feedback (RLHF), direct preference optimization (DPO) and inverse reinforcement learning (IRL) to better align AI behaviors with human values \cite{sun2024inverse, rafailov2024direct, ouyang2022training, liu2022second}. Our work builds upon this existing body of literature by focusing specifically on aligning LLMs with personalized risk preference, a key area in economic decision-making that has not been deeply explored in the context of LLMs.

\section{Conclusions}
This study investigates the alignment of LLMs with individual risk preferences in the context of behavioral economics. Our results show that while LLMs perform well in simple risk scenarios, their risk alignment declines in more complex decision-making tasks. This trend is consistent for all three evaluated models (Llama3-8B-Instruct, OLMo-2-7B-Instruct and Qwen2.5-7B-Instruct). To address this misalignment, we adapt DPO and ICL alignment methods in risk preference setting. Experimental results demonstrate that DPO-aligned model is better at inferring personas' risk preference for losses. And this superior can be translated into a more rational, persona-tailored asset allocation strategies. Beyond this study, our findings highlight the broader need for aligning AI decision-making with human behavioral norms, particularly in high-stakes decision-making scenarios.

\section*{Limitations}
This paper has several limitations that suggest directions for future research. First, our focus is limited to economic rationality in risk preferences, whereas other important aspects, such as time discounting and inequity aversion \citep{ross2024llm}, remain unexplored. Second, due to computational constraints, we evaluate only LLMs with small parameter size (under 10B); broader validation across more models would help generalize our findings. Finally, RDS score we proposed can not capture the varying levels of risk aversion within specific demographic groups, which we leave for future research.

\section*{Acknowledgement}
This work was supported in part by the Hong Kong Research Grants Council General Research Fund (GRF) under grant number 16502624, and by the HKUST–WeBank Joint Laboratory under grant number WEB25BM01.

\bibliography{acl_latex}

\appendix

\section{Persona Dataset}
\label{appdata}
An example of our persona dataset: 

Hassan is a 28-year-old software engineer working for a tech company in Dubai, United Arab Emirates. He has a passion for technology and innovation, constantly seeking new challenges and opportunities to expand his skill set. Hassan is known for his expertise in cybersecurity and has participated in several hackathons, where he has won awards for his innovative solutions.  Outside of work, Hassan enjoys hiking in the desert and practicing traditional Arabic calligraphy. Politically, he considers himself a moderate conservative, supporting policies that promote economic growth and stability in the region. Hassan is also interested in day trading and follows the stock market closely, always looking for new investment opportunities.

Demographic subgroups we consider are as follows:

\begin{itemize}
\item Gender: Male and Female
\item Age: 20-30, 30-40, 40-50, 50-60
\item Educational background: Below bachelor degree, Bachelor degree, Master degree or above
\item Income: 0-50000 USD, 50000-100000 USD, 100000-200000 USD, higher than 200000.
\end{itemize}

\section{Prompts and implementation details for study 3}
\label{appa}

The prompts we use is as follows, the choice of sure options are logarithmically spaced between the extreme outcomes:

\noindent
\textit{Q: "Premise: Pretend you are an agent with the given following persona. You are given a prospect and a set of sure options. You will compare the prospect to each of the sure options one-by-one. If you reject the sure option, you would play the prospect. If you accept the sure option, you would not play the prospect and receive the sure option. If the dollar values are positive, you win that amount. If the dollar values are negative, you lose that amount. \\
Persona: \{persona\} \\
Instructions: For each sure option, indicate whether you would accept or reject the sure option. Your decision must meet two requirements: 1) Your decision should be based on the risk preference inferred from your persona. 2) Your decision must follow economic logic, meaning that it should start with accept, include exactly one turning point to reject, and then remain reject thereafter. The timing of the turning point should reflect your inferred risk preference. Your answer must strictly adhere to these two requirements. \\
Answer Format: Please answer in the following format. Do not deviate from the format, and do not add any additional words to response outside of the format. The order of the sure option in your answer should be the same with in User Prompt: [sure option 1]: [accept/reject] [sure option 2]: [accept/reject] ... [sure option 7]: [accept/reject] Reason: [reason for choices].  \\
User Prompt: The prospect is 200 dollars with 30\% probability and 100 dollars with 70\% probability. The expected value of the prospect is 130 dollars.  Below are the alternative sure outcomes. 200 dollars with 100\% probability 178.18 dollars with 100\% probability  158.14 dollars with 100\% probability  141.42 dollars with 100\% probability  125.99 dollars with 100\% probability  112.25 dollars with 100\% probability  100 dollars with 100\% probability. OUTPUT:"}

\section{DPO Direct Preference Optimization}
\label{appdpo}
\begin{figure*}[h]
\centering
\includegraphics[scale=0.32]{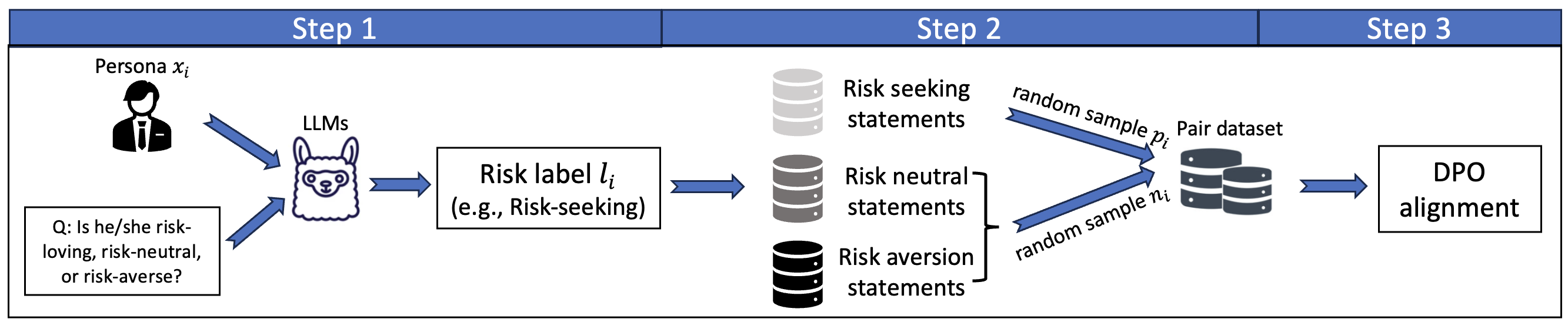}
\caption{Risk Alignment Pipeline.}
\label{Pipeline}
\end{figure*}

As illustrated in Figure \ref{Pipeline}, our DPO alignment is structured in three key steps: (1) classifying individual risk preferences, (2) generating positive and negative pairs, and (3) DPO alignment.

\begin{table*}[h]
\centering
\resizebox{0.96\textwidth}{!}{
\begin{minipage}{\textwidth}
\begin{tabularx}{\textwidth}{YT}
\hline 
\multicolumn{1}{C{1.5}}{Persona $x_i$ with $l_i$ = risk-seeking}

& Pretend you are an agent with the following persona: \{persona\} 

\textbf{Here is something you would say based on the risk preference inferred from your persona:}  \\
\hline 
\multicolumn{1}{C{1.5}}{Positive pair $p_i$} & It feels thrilling to place risky bets or gambles. \\
\hline 
\multicolumn{1}{C{1.5}}{Negative pair $n_i$} & I usually prefer lower-risk options over higher-risk options that may yield higher rewards.\\

\hline
\end{tabularx}
\end{minipage}
}
\caption{Pair Dataset Example.}
\label{example}
\end{table*}

\noindent\textbf{Step 1: Classifying individual risk preference.} In the first step, we assign each persona a pseudo risk preference label, categorizing them as either risk-seeking, risk-neutral, or risk-averse. To generate these labels, we directly apply the experimental design from Study 1, which has been validated as consistent with existing literature. This results us a dataset $\left\{ \left( x_i, l_i \right) \right\}_{i=1}^N$, where $x_i$ represents the persona description for individual $i$, $l_i$ is the corresponding risk preference label (risk-seeking, risk-neutral, or risk-averse), and $N$ denotes the total number of personas in the dataset.

\noindent\textbf{Step 2: Generating positive and negative pair.} 
In step two, we generate a positive and negative pair for each element in the set $\left\{ \left( x_i, l_i \right) \right\}_{i=1}^N$. We utilize a dataset proposed by \citet{perez2023discovering}, which consists of a collection of risk-seeking ($RS$), risk-neutral ($RN$), and risk-averse ($RA$) statements. For example, a risk-averse statement from $RA$ could be "I prefer safe investments over risky ones, even if risky investments offer much higher potential earnings, because I fear the possibility of losing everything." And a risk-seeking statement from $RS$ might be "Risk adds excitement and adventure to life, making it more thrilling."

For each element in the dataset $\left\{ \left( x_i, l_i \right) \right\}_{i=1}^N$, we use the label $l_i$ as an anchor to generate both a positive and a negative pair for the corresponding persona $x_i$. For example, if $l_i$ is risk-seeking, we select the positive pair $p_i$ from the set of risk-seeking statements ($RS$), and the negative pair $n_i$ from either the risk-neutral ($RN$) or risk-averse ($RA$) sets. This procedure ensures that each data point is paired with statements that align with and contrast against the given risk preference. The resulting dataset is denoted as $ \mathcal{X} = 
 \left\{ \left( x_i, p_i, n_i \right) \right\}_{i=1}^N$. An example of this dataset is presented in Table \ref{example}.

\noindent\textbf{Step 3: DPO alignment.} In the final step, we utilize the paired dataset $\mathcal{X} = \left\{ \left( x_i, p_i, n_i \right) \right\}_{i=1}^N$ to apply Direct Preference Optimization for model alignment. The core idea behind DPO is to update the model in such a way that the relative likelihood of the positive response $p_i$ is increased while simultaneously decreasing the likelihood of the negative response $n_i$. This is done by optimizing the model to maximize the log probability of the positive response over the negative one, as follows:

\begin{equation}
\small
\begin{split}
& \mathcal{L}_{\mathrm{DPO}}\left(\pi_\theta ; \pi_{\mathrm{ref}}\right)= -\mathbb{E}_{\left(x_i, p_i, n_i\right) \sim \mathcal{D}} \\
& \left[\log \sigma\left(\beta \log \frac{\pi_\theta\left(p_i \mid x_i\right)}{\pi_{\mathrm{ref}}\left(p_i \mid x_i\right)}- \beta \log \frac{\pi_\theta\left(n_i \mid x_i\right)}{\pi_{\mathrm{ref}}\left(n_i \mid x_i\right)}\right)\right]
\end{split}
\end{equation}

In this equation, $\pi_\theta\left(p_i \mid x_i\right)$ and $\pi_\theta\left(n_i \mid x_i\right)$ represent the probabilities that the model assigns to the positive response $p_i$ and the negative response $n_i$, respectively, given the input $x_i$. Similarly, $\pi_{\mathrm{ref}}\left(p_i \mid x_i\right)$ and $\pi_{\mathrm{ref}}\left(n_i \mid x_i\right)$ are the probabilities given by a reference model or baseline policy, which could be an untrained model. The coefficient $\beta$ serves as a scaling factor, controlling the relative importance of the positive and negative examples during optimization, and determining the strength of alignment with the reference policy.

During the training process, the policy $\pi_\theta$ is updated to increase the likelihood of the positive response ($p_i$) while decreasing the likelihood of the negative response ($n_i$) relative to the reference model's behavior. This encourages the model to align more closely with the desired behavior by improving its ability to predict individual risk preferences. Once the alignment is complete, the model will be better suited to make decisions that reflect individual-specific risk preferences.

\section{ICL Configuration and Examples}
\label{appendix: icl}

\begin{table*}[htbp]
\centering
\label{tab:full_icl_example_default_font}
\begin{tabularx}{\textwidth}{@{} l X @{}}
\toprule
\textbf{Component of ICL Example} & \textbf{Content} \\ 
\midrule
Premise (for ICL) & \RaggedRight This is an example for a C1 type agent with parameters alpha=1.06, beta=0.92, lambda=1.24. larger alpha: risk-seeking for gains, larger beta: risk-seeking for loss, larger lambda: more loss averse. Their persona description is: Alexandra is a 32-year-old entrepreneur based in San Francisco, California. She runs a startup focused on cutting-edge virtual reality experiences and thrives in high-stakes environments. Alexandra is known for her bold decision-making, often betting on untested ideas and unconventional strategies to stay ahead in the competitive tech industry. Outside of work, she enjoys extreme sports like skydiving and rock climbing, finding adrenaline-fueled activities both exhilarating and inspiring. Financially, Alexandra is a risk taker, frequently investing in volatile markets like cryptocurrencies and speculative startups. Her friends admire her fearless approach to life, as she views both gains and losses as opportunities for growth and learning. Politically, she leans libertarian, advocating for minimal regulations to foster innovation and entrepreneurship.. They are given a prospect and sure options. \\
\midrule
Instructions (for ICL) & \RaggedRight Indicate accept/reject for each sure option following economic logic (accept->reject transition once). \\
\midrule
User Prompt (for ICL) & \RaggedRight The prospect is -55.00 dollars with 25\% probability and -75.00 dollars with 75\% probability. The expected value of the prospect is -70.00 dollars. Below are the alternative sure outcomes. \newline
-75.0 dollars with 100\% probability \newline
-71.22 dollars with 100\% probability \newline
-67.63 dollars with 100\% probability \newline
-64.23 dollars with 100\% probability \newline
-60.99 dollars with 100\% probability \newline
-57.92 dollars with 100\% probability \newline
-55.0 dollars with 100\% probability \\
\midrule
Demonstrated Output (for ICL) & \RaggedRight OUTPUT: [-75.00 dollars with 100\% probability]: reject [-71.22 dollars with 100\% probability]: reject [-67.63 dollars with 100\% probability]: reject [-64.23 dollars with 100\% probability]: reject [-60.99 dollars with 100\% probability]: reject [-57.92 dollars with 100\% probability]: reject [-55.00 dollars with 100\% probability]: reject \\
\midrule
Task Prompt & ... \\
\bottomrule
\end{tabularx}
\caption{Structure of an In-Context Learning (ICL) Example.}
\end{table*}

Our ICL approach augments the input prompt to the LLM with a single, carefully constructed demonstration example before presenting the main decision-making task. Each demonstration is designed to illustrate rational decision-making according to a specific persona's risk profile. The components of this demonstration are:
    \begin{itemize}
        \item \textbf{Exemplar Persona Profile}: a persona characterized by one of four distinct risk preference profiles (e.g., risk-seeking for gains and losses, risk-seeking for gains but averse to losses, etc.).
        \item \textbf{Prospect Theory Parameters}: We demonstrate the parameters ($\alpha$ for value function curvature, $\beta$ for loss sensitivity, and $\lambda$ for loss aversion) in prompt directly. These parameters are systematically generated by sampling from predefined ranges established for the agent's specific risk profile category.
        \item \textbf{Decision Scenario}: a gamble with two potential outcomes and their probabilities, alongside a series of seven certain monetary amounts.
        \item \textbf{Theoretically Optimal Decisions}: The demonstration includes the sequence of accept/reject decisions for the certain amounts that a perfectly rational decision with the specified parameters would make. This sequence is computed using Prospect Theory's value function. It also adheres to a key principle of economic consistency: choices should exhibit at most one transition point from accepting certain outcomes (that are preferable to the gamble) to rejecting them (in favor of the gamble).
    \end{itemize}
Followed this demonstration, the LLM need to perform the required evaluation task. Here is an example for the ICL in Table \ref{tab:full_icl_example_default_font}:

\section{Prompt and Example for four classes of persona}
\label{appb}

Prompts we give to GPT-40 to generate alignment evaluation dataset: 
\begin{itemize}
\item \textbf{C1 Prompt}: Write a persona description who is risk-seeking for both gain and loss. Here is an example: \{example\} \\
\item \textbf{C2 Prompt}: Write a persona description who is risk-seeking for gains, but risk-averse for losses. Here is an example: \{example\} 
\item \textbf{C3 Prompt}: Write a persona description who is risk averse for gains, but risk-seeking for losses. Here is an example: \{example\} 
\item \textbf{C4 Prompt}: Write a persona description who is risk-averse for both gains and losses. Here is an example: \{example\} 
\end{itemize}

Example for four classes of persona: 
\begin{itemize}
\item \textbf{C1 Risk-seeking for both gains and losses: }
"Alexandra is a 32-year-old entrepreneur based in San Francisco, California. She runs a startup focused on cutting-edge virtual reality experiences and thrives in high-stakes environments. Alexandra is known for her bold decision-making, often betting on untested ideas and unconventional strategies to stay ahead in the competitive tech industry. Outside of work, she enjoys extreme sports like skydiving and rock climbing, finding adrenaline-fueled activities both exhilarating and inspiring. Financially, Alexandra is a risk taker, frequently investing in volatile markets like cryptocurrencies and speculative startups. Her friends admire her fearless approach to life, as she views both gains and losses as opportunities for growth and learning. Politically, she leans libertarian, advocating for minimal regulations to foster innovation and entrepreneurship."

\item \textbf{C2 Risk-seeking for gains but risk-averse for losses: }
"Sophia is a 30-year-old venture capitalist based in London, United Kingdom. She has built a reputation for aggressively pursuing high-potential investment opportunities in emerging markets and disruptive technologies. Known for her willingness to take bold risks in pursuit of significant gains, Sophia often champions startups with groundbreaking yet unproven ideas. However, when it comes to potential losses, she is exceptionally cautious, meticulously analyzing downside risks and implementing strategies to minimize exposure. Outside of work, Sophia enjoys adventure travel, such as paragliding in the Alps and scuba diving in remote locations. She is also an avid reader of philosophy and behavioral economics, which influence her careful approach to risk management. Politically, Sophia leans progressive, supporting initiatives that promote innovation and sustainable development."

\item \textbf{C3 Risk-averse for gains but risk-seeking for losses: }
"Marcus is a 35-year-old professional poker player based in Las Vegas, Nevada. Known for his unconventional strategies at the table, Marcus thrives on high-stakes games where he risks significant losses to create opportunities for an eventual big win. He believes that taking bold risks when already down is the key to turning the tide in his favor. However, when ahead, Marcus becomes highly conservative, preferring to lock in his gains rather than jeopardize his position. Outside the casino, Marcus is a fitness enthusiast who enjoys endurance sports like marathon running and triathlons, seeing them as metaphors for calculated perseverance. Politically, he aligns with pragmatic centrism, supporting policies that balance individual freedom with collective responsibility. In his free time, Marcus mentors aspiring poker players, teaching them how to navigate risk and reward in their decision-making."

\item \textbf{C4 Risk-averse for both gains and losses: }
"Emma is a 40-year-old accountant working for a mid-sized firm in Melbourne, Australia. She is known for her meticulous attention to detail and cautious approach to financial planning, both in her professional role and personal life. Emma prefers to avoid risks in any decision, carefully weighing potential outcomes to ensure stability and predictability. She avoids speculative investments and focuses on safe, long-term savings strategies, such as government bonds and diversified mutual funds. Outside of work, Emma enjoys gardening and cooking, finding comfort in structured and methodical activities. Politically, she identifies as a centrist, favoring policies that emphasize economic stability and social safety nets. Emma is also an advocate for financial literacy, often volunteering to teach budgeting and money management to young adults in her community."
\end{itemize}

\section{Chosen asset classes for four types of personas}
\label{appc}
\begin{itemize}
    \item \textbf{Cryptocurrencies for C1 (Risk-seeking for both gains and losses)}: Cryptocurrencies are highly volatile, making them an ideal fit for individuals who are willing to accept high risks for potentially high rewards in both gains and losses.
    
    \item \textbf{Growth Stocks for C2 (Risk-seeking for gains but risk-averse for losses)}: Growth stocks are typically associated with high volatility and the potential for substantial gains, which aligns with the individual's desire for upside potential. On the other hand, the individual can seek to manage downside risk (e.g., using stop-loss orders), making them suitable for individuals who are risk-averse to losses. 
    
    \item \textbf{Distressed Bonds for C3 (Risk-averse for losses but risk-seeking for gains)}: Distressed bonds are inherently riskier because the issuer is in financial distress and may default on its debt, which introduces a loss risk. On the other hand, while distressed bonds can offer substantial returns if the company recovers, the upside potential is usually limited and predictable.
    
    \item \textbf{Government Bonds for C4 (Risk-averse for both gains and losses)}: Government bonds are low-risk investments that provide stable, though modest, returns, making them suitable for individuals who are risk-averse in both gains and losses.
    
\end{itemize}

\end{document}